\documentclass[runningheads]{llncs}
\usepackage[T1]{fontenc}
\usepackage{graphicx}
\usepackage{hyperref}
\usepackage{amssymb}
\usepackage{amsmath}

\begin{document}
\title{Super Images - A New 2D Perspective on 3D Medical Imaging Analysis}
\titlerunning{Super Images}
%
\author{Ikboljon Sobirov\and
Numan Saeed \and
Mohammad Yaqub}
\institute{Mohamed bin Zayed University of Artificial Intelligence, Abu Dhabi, UAE
\email{\{ikboljon.sobirov, numan.saeed, mohammad.yaqub\}@mbzuai.ac.ae}}
\maketitle              
\begin{abstract}
In medical imaging analysis, deep learning has shown promising results. We frequently rely on volumetric data to segment medical images, necessitating the use of 3D architectures, which are commended for their capacity to capture interslice context. However, because of the 3D convolutions, max pooling, up-convolutions, and other operations utilized in these networks, these architectures are often more inefficient in terms of time and computation than their 2D equivalents. Furthermore, there are few 3D pretrained model weights, and pretraining is often difficult. We present a simple yet effective 2D method to handle 3D data while efficiently embedding the 3D knowledge during training. We propose transforming volumetric data into 2D super images and segmenting with 2D networks to solve these challenges. Our method generates a super-resolution image by stitching slices side by side in the 3D image. We expect deep neural networks to capture and learn these properties spatially despite losing depth information. This work aims to present a novel perspective when dealing with volumetric data, and we test the hypothesis using CNN and ViT networks as well as self-supervised pretraining. While attaining equal, if not superior, results to 3D networks utilizing only 2D counterparts, the model complexity is reduced by around threefold. Because volumetric data is relatively scarce, we anticipate that our approach will entice more studies, particularly in medical imaging analysis.

\end{abstract}
\section{Introduction}
3D medical imaging modalities, such as computed tomography (CT), magnetic resonance imaging (MRI), and positron emission tomography (PET), are used extensively in clinical practice. Doctors rely on them to understand the volumetric information (depth of tumor, etc.) to perform their diagnosis more accurately. As such, a significant number of the developed data-driven techniques for volumetric medical images process the data in 3D using 3D deep neural networks. While such an approach produces promising results in the research community, its applications are limited in clinical practice. Although it may hinder the accuracy of the solution, 2D approaches have benefits such as faster and more cost-effective performance. Methods which could utilize the benefits of 2D slice-wise and 3D processing are of great importance.

The real-life medical applications for volumetric data analysis via 3D networks are still limited due to several reasons, one of which is the high complexity of the models, especially when working with volumetric data. Hence, the deployment of such DL models in medical practices becomes difficult. Recent works~\cite{transfuse,feng} mainly focus on increasing the performance by a small margin, which in turn increases the model sizes in parameters and FLOPs. We argue that the reduction of model sizes while keeping similar performance encourages for easier deployment of these models in medical practices. 

It is true that performing segmentation using the 3D data directly is praised to produce better results. Several authors~\cite{3dcompact,3dunet,3dvnet} support using 3D datasets primarily because 3D networks can capture depth information that their 2D counterpart lacks and claim that this information is crucial to model learning. Another argument is that the nature of 3D data is closer to real life, which is why 3D models ought to perform better~\cite{3dcompact,3dvnet}. On the downside, instead of 2D, 3D convolutions, max pooling and up-convolutions are applied during model learning, thus requiring much more computation power and training/inference time~\cite{2dguy,super}. 

On the other hand, those who argue that 2D should still be in heavy use reinforce their claim that utilizing 2D images is more cost- and time-effective and offers more options to apply transfer learning~\cite{super,2dnode,2dreinvent}. Transfer learning, with weights pretrained on large-scale datasets, such as ImageNet~\cite{imagenet}, can be considerably beneficial to model learning, especially when medical datasets are scarce. Another advantage of using 2D images is that 3D can be easily converted to multiple 2D slices, generating a larger scale set than relying on a limited number of 3D counterparts. Another vivid upside is that there are numerous 2D architectures available for the encoder of U-Net~\cite{3dunet} like models~\cite{2dguy2}. This makes 2D models easier to customize and adjust to the need of the problem at hand.

\begin{figure}[ht!]
\centerline{\includegraphics[width=\columnwidth]{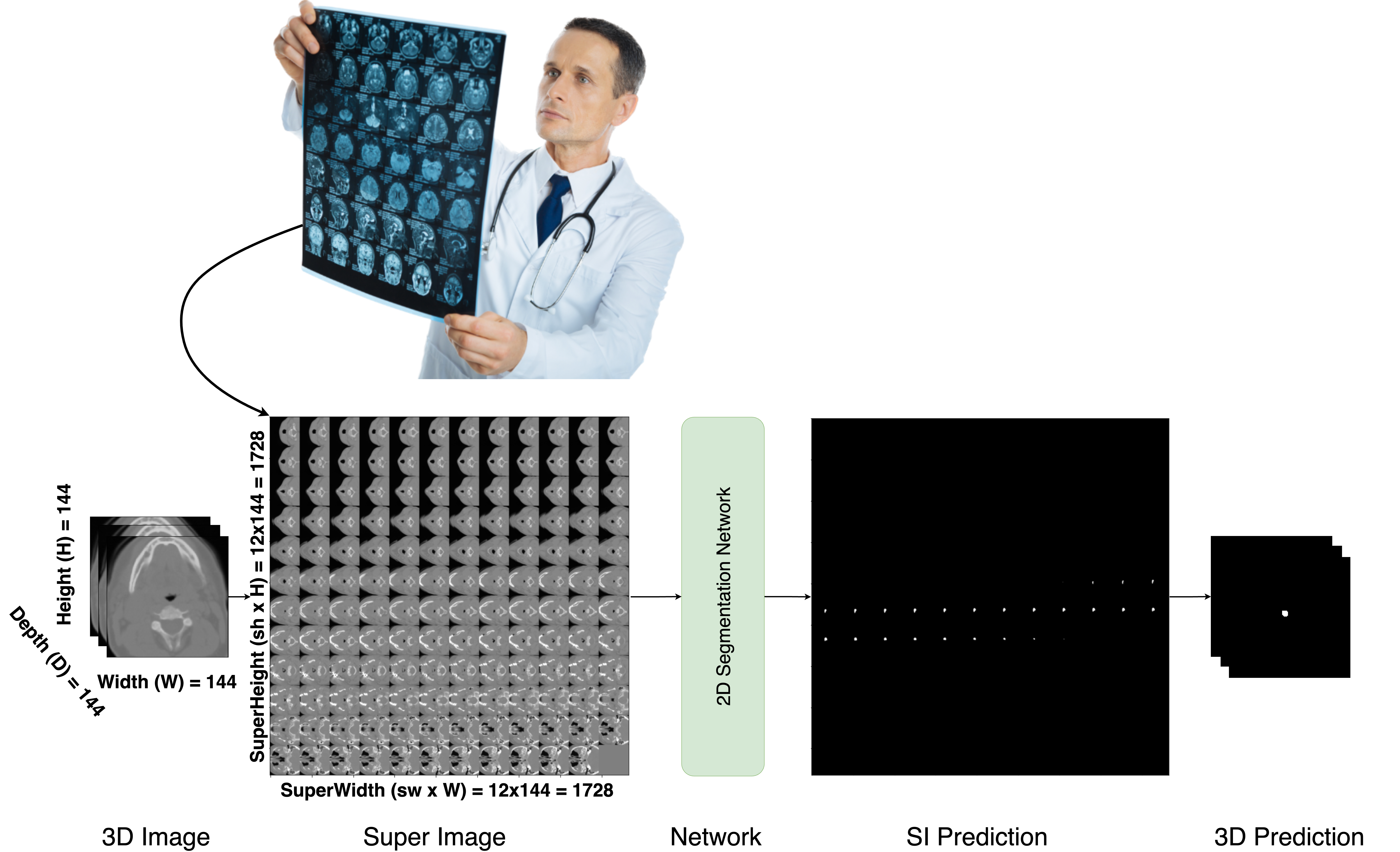}}

\caption{(Top) The figure shows the intuition behind using super images. Similar to how clinicians examine the scans from a bird's-eye view in a slice-wise comparative approach, we generate SIs side by side such that DL models can analyze the images in the same fashion. Image: Adobe Stock~\cite{yakobchuk}. (Bottom) The figure shows the construction of super images from volumetric data. We rearrange the depth dimension by assembling the slices to generate the super image. It is then fed to a 2D segmentation network. The model yields the prediction mask, which is then rearranged back to the original shape. Note that the volumetric prediction mask shows a tumor region for visualization purposes.
}
\label{super_image}
\end{figure}

Considering the abovesaid advantages of the 2D approach, this paper introduces a new perspective on using volumetric data in a 2D fashion. The intuition behind the approach is visualized in Figure~\ref{super_image} (top), where a clinician is examining the scan from a bird's-eye view for slice-wise comparative analysis. We generate 2D super images (SIs) from 3D input by stacking the depth information (i.e., slices) side-by-side, and we train a 2D network for the same task. A similar notion was proposed by~\cite{super} on natural videos; unlike them, this concept is newly introduced to the medical field and volumetric data in particular. This novel approach in the segmentation task can achieve comparable results to a 3D model counterpart and reduce the model complexity by around threefold. The main contributions of our paper are:

\begin{itemize}
    \item We introduce a new perspective to dealing with biomedical volumetric data by casting them into super images and training 2D models with them.
    \item We empirically show that pretraining techniques can easily boost the performance of the models that use super images.
    \item We validate our method on different CT, PET, and MRI datasets to show the effectiveness of our approach.
    \item The approach achieves comparable results to a 3D model counterpart and reduce the model complexity by around threefold.
\end{itemize}

\section{Methodology}

The proposed approach is relatively simple to implement yet effective in training. In brief, volumetric data are converted to SIs, a 2D network of choice is trained on them, and the model outputs are cast back to the original dimensions. More details are provided below.

\subsection{Super Image Generation}
A 3D volume can provide features from the depth information for the model to learn since they use 3-dimensional kernels. Still, we expect these characteristics to be detectable and learnable in 2D SIs by well-designed deep neural networks. With that in mind, we generate SIs from volumetric data by taking slices and stitching them together side by side in order, as shown in Figure~\ref{super_image}(bottom). Given a 3D image $x_{inp}\in \mathbb{R}^{H\times W\times D\times C}$, where $H$ is the height, $W$ is the width, $D$ is the depth, and $C$ is the number of channels, the depth dimension is rearranged. The resulting image $s_{inp}\in \mathbb{R}^{\hat{H}\times\hat{W}\times C}$ is now 2D, where $\hat{H} = H\times sh$, and $\hat{W} = W\times sw$; $sh$ and $sw$ represent the degree by which the height and width should be rearranged respectively to generate a grid size of $sh\times sw$. As a demonstration, the size of $144\times144\times144\times2$ (2 for CT and PET), having $144$ as the depth, can be considered with $sh$ of 12 and $sw$ of 12, thus generating the SI in the dimensions of $1728\times1728\times2$. 

2D U-Net (or any other 2D segmentation model, for that matter) is trained on these SIs to perform the segmentation. The model output in the dimensions of $s_{out}\in \mathbb{R}^{\hat{H}\times\hat{W}\times C}$ is rearranged back to the original data dimensions of $x_{out}\in \mathbb{R}^{H\times W\times D\times C}$ as depicted in Figure~\ref{super_image} (bottom). Note that the predicted mask on the final 3D data is shown with a large tumor size only for visualization purposes.

\subsection{Experiments}
The approach of using SIs in a 2D fashion is tested with several architectures to see the performance discrepancies and model complexities between various models. All models were compared in 2D with SIs and 3D with the volumetric data. The first model was the vanilla U-Net~\cite{3dunet}, which was the foundational medical image segmentation model. The second model was modified U-Net with a squeeze and excitation normalization in the encoder and decoder~\cite{senet}, which showed better performance (here dubbed as SE-norm U-Net). We also experimented with vision transformer architectures; specifically, we studied the behavior of Swin UNETR~\cite{hatamizadeh2022swin} both in 2D with SIs and 3D with the volumetric data. The implementation of Swin UNETR in both 2D and 3D is relatively easy, and the model is praised as one of the latest and best-performing ViT networks.

On top of training from scratch, we performed SSL-based pretraining for SIs. Inpainting (i.e., masking) and jigsaw puzzle approaches were experimented with as objective tasks where the perturbed image was reconstructed during pretraining. During finetuning, the models were initialized with the pretrained weights rather than the random initialization.

\begin{figure}[t!]
\centering
\begin{tabular}{ccc}
{\includegraphics[width=0.3\columnwidth]{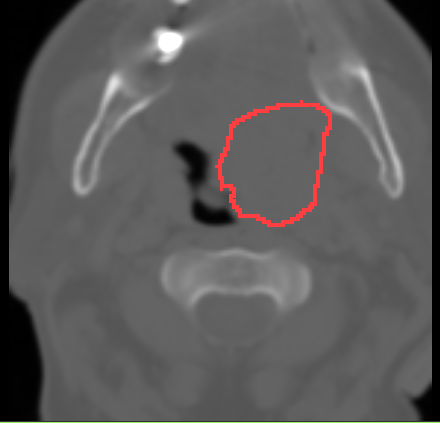}}&
{\includegraphics[width=0.3\columnwidth]{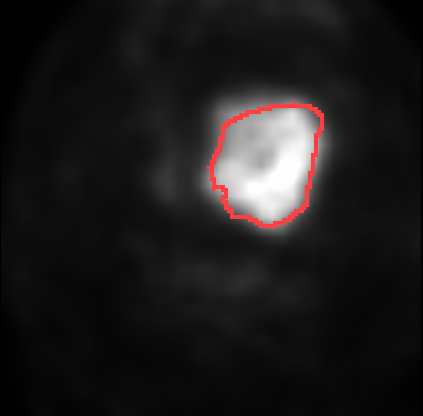}}&
{\includegraphics[width=0.3\columnwidth]{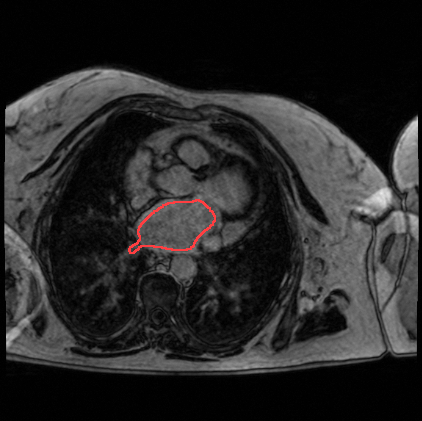}}
\\
(a)&(b)&(c)
\end{tabular}
\caption{The figure shows sample slices from the two datasets. (a) and (b) depict CT and PET scans with red regions highlighting tumors from the HECKTOR dataset, respectively; and (c) shows an MRI slice with the red region corresponding to the atrium from the Atria segmentation dataset.}
\label{sample}
\end{figure}

\subsection{Datasets \& Preprocessing}
To validate our new approach, we experimented with two different datasets: head and neck tumor segmentation and outcome prediction (HECKTOR) challenge~\cite{hecktor} and atrial segmentation challenge~\cite{asc} datasets. Several preprocessing techniques are performed on both datasets accordingly.

\paragraph{Head and Neck Tumor:} HECKTOR dataset comprises 224 CT and PET scans of patients with head and neck tumors for the training set (i.e., the dataset is available online\footnote[1]{aicrowd.com/challenges/miccai-2021-hecktor}). Bounding box information comes with the dataset for localization of the tumor region, which was used to crop the scans and the mask down to the size of $144\times144\times144mm^3$ with consistency between the scans. Sample CT and PET slices are depicted in Figure~\ref{sample} (a) and (b), respectively, where the red area corresponds to the tumor region. Since the challenge organizers provide the bounding box information, the tumor is within the cropped region, and mappings between both modalities and mask are accurate. Further preprocessing techniques were re-sampling the data to have isotropic voxel spacing ($1\times1\times1mm^3$) and the intensity normalization of both CT and PET data. CT scans were clipped in the range of (-1024, 1024) and normalized to (-1, 1), and Z-normalization was used for PET scans.

\paragraph{Atria:} Atrial segmentation challenge dataset includes 100 3D gadolinium contrast (GE) MRIs for the training set\footnote[2]{atriaseg2018.cardiacatlas.org/data}. Figure~\ref{sample} (c) shows an MRI sample with a red line delineating the atrial region. The scans are of different dimensions and thus were resized to the same size of $512\times512\times88mm^3$. Similarly, intensity normalization was applied to the scans. 

No further data augmentations were applied on either dataset unless reported otherwise. The testing set ground truth is inaccessible in both datasets; therefore, they are not used, and instead, $k$-fold cross validation was utilized for all the experiments. 

We purposely chose two datasets of different modalities (i.e., CT and PET in one and MRI in another) to test the hypothesis for generalizability. Moreover, the tasks are the tumor segmentation in one dataset and atria in the other. This shows that the idea is not limited to a specific tissue type with specific characteristics. 

\begin{table}[t!]
\centering
\caption{Mean values of DSC, precision, recall and HD95, and the number of parameters and FLOPs of different models on the validation set from 5-fold cross-validation are reported. Dim. correspond to dimensionality (either 3D volumetric or 2D SI). LK stands for large kernel.}
\begin{tabular}{ cc }   
    \centering

    {\begin{tabular}{l|ccccccc}
    Models & Dim. & DSC & Precision & Recall & HD95 & Params (M) & FLOPs (G) \\
    \hline
    U-Net & 3D & 0.728 & 0.734 & 0.783 & 3.035 & 3.61 & 518.61 \\
    U-Net & 2D SI & 0.725 & 0.743 & 0.772 & 4.689 & 1.21 & 319.98 \\
    \hline
    SE-norm U-Net & 3D & 0.747 & 0.766 & 0.789 & 3.638 & 21.75 & 642.6\\
    SE-norm U-Net & 2D SI & 0.737 & 0.767 & 0.764 & 5.185 & 8.51 & 493.87 \\
    SE-norm LK & 2D SI & 0.739 & 0.757 & 0.780 & 4.170 & 28.09 & 1181.84 \\
    \hline
    Swin UNETR & 3D & 0.729 & 0.744 & 0.775 & 3.110 & 15.7 & 280.53 \\
    Swin UNETR & 2D SI & 0.732 & 0.753 & 0.774 & 5.176 & 6.3 & 214.86 \\
    
\end{tabular}}

\end{tabular}
\label{tab:main_results}
\end{table}

\section{Experimental Setup}
We used a single NVIDIA RTX A6000 GPU for our experiments, and the implementation was done using the PyTorch library. We ran all the experiments for 100 epochs (both random initialization and finetuning). Pretraining was performed for 300 epochs. An AdamW optimizer with the initial learning rate of 0.001 and weight decay of 1e-5 was used, and a cosine annealing schedule that starts with the initial learning rate, decreasing it to the base learning rate of 1e-5 and resetting it after every 25 epochs were chosen to control the learning rate. The batch size was set to 2 for the HECKTOR dataset.  The dice similarity coefficient (DSC) was chosen as the primary evaluation metric, and additional 95\% Hausdorff distance (HD95), precision, and recall were also calculated. Since we claim that the 2D approach is less costly, the number of parameters (in M) and FLOPs (in G) are also provided for all the experiments/models. $k$-fold cross-validation was utilized for all the experiments, and their mean values are listed in Section~\ref{sec:results}.

\section{Results}
\label{sec:results}
We used two different datasets to verify this new approach to dealing with volumetric data. We used various models for the HEKCTOR dataset to see the approach's applicability. The atrial segmentation dataset was also used as additional support to showcase the generalizability of the approach.


Table~\ref{tab:main_results} shows the mean results of 5-fold cross-validation for the HECKTOR dataset for different models. Dim. column indicates the dimensionality of the model used for that experiment, with 2D SI meaning a 2D model trained on SIs. The vanilla U-Net models on 3D and 2D SI achieved similar DSC of 0.728 and 0.725, respectively, whereas the number of parameters for the 2D model is three times less than that of the 3D model. Similar values can be observed for the other metrics for both approaches too.

\begin{table}[t!]
\centering
\caption{The table shows the results of ablation study on the grid size of the super images. The HECKTOR dataset scans were cropped around the tumor region to form a small size of $80\times80\times48$ for a quick experimentation. The SIs were arranged in different ways, i.e. the $sw$ and $sh$ were chosen with different arrangements to see how it would effect the model learning. The results are the means of 5-fold cross validation. The results indicate that the more square-like formation of SIs perform better than elongated rectangular arrangements.}\label{hecktor_results}
\setlength{\tabcolsep}{6pt}
\begin{tabular}{l|l|l|l|l|l|l}
\hline
Model &  Image Size & \textit{sh} & \textit{sw} & DSC & Precision & Recall \\
\hline
3D U-Net & $80\times80\times48$ & - & - & 0.779 & 0.787 & 0.822 \\
2D U-Net & $640\times480$ & 8 & 6 & 0.778 & 0.799 & 0.810 \\
2D U-Net & $480\times640$ & 6 & 8 & 0.777 & 0.793 & 0.816 \\
2D U-Net & $960\times320$ & 12 & 4 & 0.770 & 0.809 & 0.801 \\
2D U-Net & $320\times960$ & 4 & 12 & 0.759 & 0.790 & 0.797 \\
2D U-Net & $1920\times160$ & 24 & 2 & 0.744 & 0.765 & 0.809 \\
2D U-Net & $160\times1920$ & 2 & 24 & 0.762 & 0.779 & 0.809 \\

\end{tabular}
\end{table}

With the SE-norm U-Net, the performance increased across the metrics. Again the 3D training against the 2D with SIs reached similar DSC of 0.747 and 0.737, respectively. The model parameters for 3D, however, is 21.75M in contrast to 8.51M with the 2D approach. Before moving to the ViT models, we additionally experimented with a larger kernel model for the SI approach with the 2D network. We increased the kernel size to 7 (as opposed to 3 in the vanilla approach) in the training, increasing the model complexity (28.09M), but this did not seem to show too much of an improvement in DSC. Precision, recall and HD95 values were similar in the three experiments. 

Swin UNETR performance for the two approaches were also similar. The 3D network training with volumetric data achieved DSC of 0.729, with a model complexity of 15.7M params, whereas the 2D SIs approach reached a slightly better DSC of 0.732 with only 6.3M params. Although precision and recall were similar for the two experiments, HD95 for the SI-based learning was higher with Swin UNETR model. This is hypothesized to be caused by the nature of HD95 calculation on the 2D plane against 3D plane. 

Further ablation studies were conducted to study the behavior of the image arrangements as listed in Table~\ref{hecktor_results}. The HECKTOR dataset scans were cropped around the tumor region to form a small size of $80\times80\times48$ for a quick experimentation. The SIs were arranged in different ways, i.e. the $sw$ and $sh$ were chosen with different arrangements to see how it would effect the model learning. The results shown are the means of 5-fold cross validation. The results indicate that the more square-like formation of SIs perform better than elongated rectangular arrangements.

Finally, SSL pretraining was applied for model initialization for the HECKTOR dataset. Masking (i.e. inpainting) and jigsaw puzzle techniques were selected as the pretraining tasks. During pretraining, the models were trained for 300 epochs for the reconstruction. During finetuning they were trained for 100 epochs. Masking and jigsaw with the SE-norm U-Net showed an improvement in DSC, reaching 0.741 and 0.738, respectively. Swin UNETR was experimented with the masking approach, ending up with a similar performance in DSC. Interesting observation here is that all the models achieved better HD95 values with the pretraining.

\begin{table}[t!]
\centering
\caption{The table shows the results of vanilla 3D U-Net (comparison target) to SI-based 2D U-Net on the atrial segmentation dataset. The results are the mean of 4-fold cross validation. PT stands for 2D U-Net pretrained on ImageNet1k, and A stands for augmentations.}\label{asc_results}
\begin{tabular}{l|ccccccc}
Model & Dim. &  Image Size & \textit{sh} & \textit{sw} & DSC & Precision & Recall \\
\hline
U-Net & 3D & $512\times512\times88$ & - & - & 0.893 & 0.898 & 0.894 \\
U-Net & 2D SI & $5632\times4096$ & 11 & 8 & 0.812 & 0.902 & 0.785 \\
\hline
U-Net & 2D SI & $4096\times4096$ & 8 & 8 & 0.851 & 0.913 & 0.822 \\
PT & 2D SI & $4096\times4096$ & 8 & 8 & 0.895 & 0.872 & 0.878 \\
PT\&A & 2D SI & $4096\times4096$ & 8 & 8 & 0.901 & 0.919 & 0.890 \\

\end{tabular}
\end{table}

In the atrial segmentation task, because the dataset contains only 100 scans, we used 4-fold cross-validation, leaving more to the validation set for better generalizability. In this set of experiments listed in Table~\ref{asc_results}, there are two separate settings: (i) U-Net comparison and (ii) experimental image preprocessing for SIs. The first setting is to repeat the performance generalizability in a different task, and the second setting now focuses on how preprocessing techniques can easily be used and can boost the performance of the SI-based model.

For the first setting of the comparison of the networks, the images with the size of $512\times512\times88$ were used. The DSC of 0.893, the precision of 0.898, and the recall of 0.894 were achieved with the 3D U-Net. The SI generation with this size used the grid layout of $11\times8$. From ablation studies provided in Table~\ref{hecktor_results}, we can see that a better arrangement for SIs is having a square-like grid of SIs rather than elongated rectangles. A grid size of $11\times8$ was not the most favorable combination for generating SIs since its aspect ratio is high; therefore, it could reach only 0.812 DSC, 0.902 precision, and 0.785 recall values.

In the second setting, the images were preprocessed. This component was performed to show that preprocessing techniques are easily applicable and that they can help the model improve. We started with a one-on-one aspect ratio for the SIs, having 64 slices. Simple preprocessing such as this boosted the DSC score of the 2D model to 0.851. In the next step, 2D U-Net was initialized with ImageNet1k pretrained weights, pushing the DSC to 0.895, which is a 0.044 DSC jump from the base model. In the last experiment, this pretrained model was used with several sets of augmentations to see how far the model could go using a simple 2D U-Net. The augmentations were random flip, random affine, random elastic deformation, random anisotropy, and random gamma, and they are specific only for this experiment. This aggressive augmentation pushed the DSC to 0.901, which provides a substantial increase from the baseline that could reach 0.812 DSC. This shows that small preprocessing techniques are highly useful for the SI-based network to learn a better representation.

\subsection{Qualitative Results}
\begin{figure}[t!]
\centering
\begin{tabular}{ccc}
{\includegraphics[width=0.4\columnwidth]{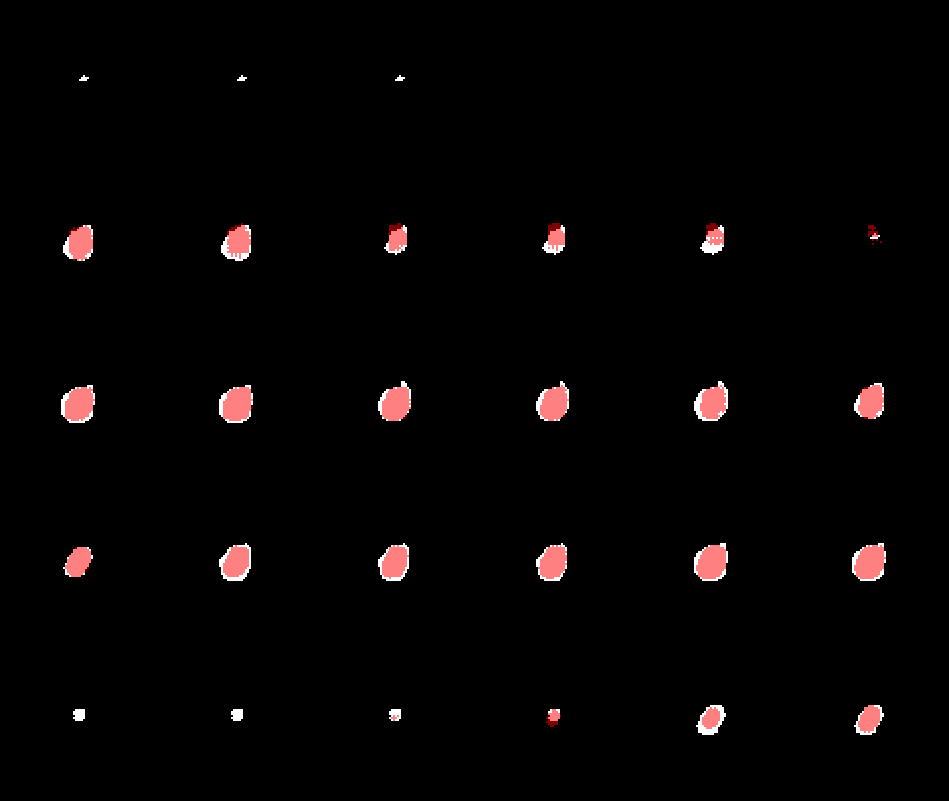}}&
{\includegraphics[width=0.4\columnwidth]{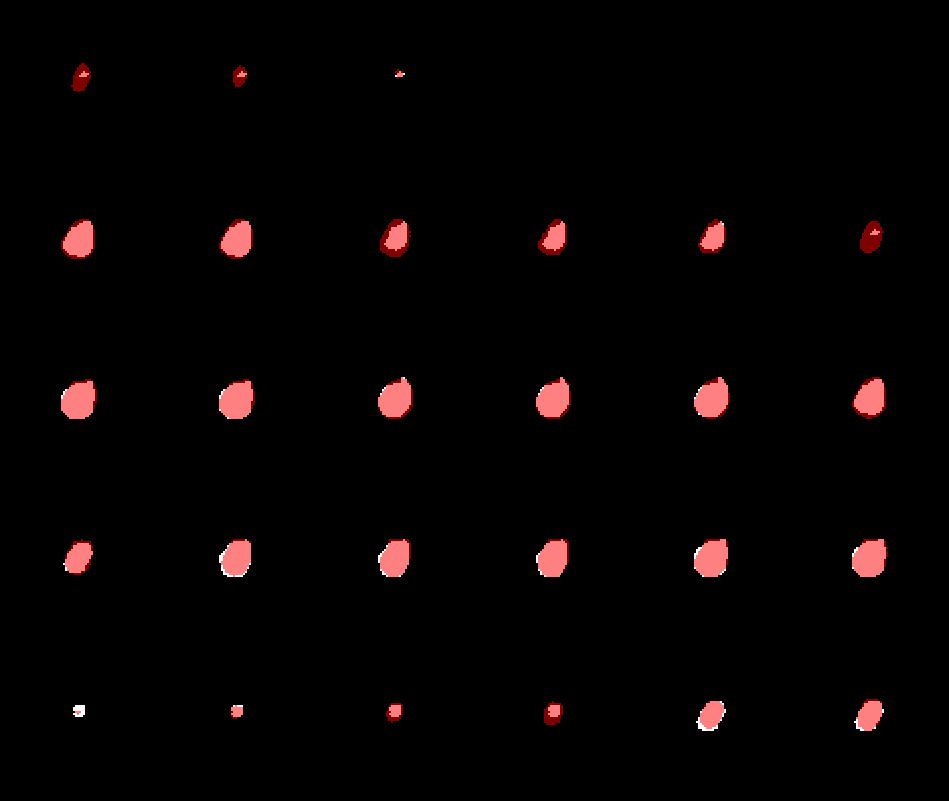}}
\\
(a)&(b)
\end{tabular}
\caption{The figure shows qualitative results on 3D U-Net (on volume) and 2D U-Net (on SI) segmentation results on a HECKTOR dataset sample, respectively. Note that the inference is on the cropped size of $80\times80\times48$. White is the ground truth and red represents the prediction mask. Note that 3D U-Net results were cast to an SI form after its prediction for full-view comparison. 3D network results (left) are much stricter than the 2D results (right), whereas the 2D model allows more oversegmentation, especially in the small size tumor regions.}
\label{qual}
\end{figure}

We studied the performance comparison in terms of qualitative analysis as well. Figure~\ref{qual} shows a sample slice from the HECKTOR dataset for 3D U-Net and 2D U-Net, respectively. Note that here we performed inference on the cropped size of $80\times80\times48$. White and red represent ground truth and model prediction, respectively. The prediction from the 3D U-Net was cast to super image formation to compare the full-view for both model predictions. In most of the visualized scans, the 3D network produces results that are much closer to the ground truth but are generally undersegmented; as such, beginning and ending slices (i.e. tiny tumor regions) are missed. The 2D network is more flexible with that, generally oversegmenting slightly more so as to capture even the tiny regions. 

\begin{table}[t!]
\centering
\caption{The table shows the results of different models that are initialized with the pretrained weights and finetuned for the task. Mean values of DSC, precision, recall, and HD95 of 5-fold cross validation are reported.}
\begin{tabular}{ cc }   
    \centering

    {\begin{tabular}{l|ccccc}
    Models & SSL & DSC & Precision & Recall & HD95 \\
    \hline
    SE-norm U-Net & Masking & 0.741 & 0.764 & 0.773 & 4.708 \\
    SE-norm U-Net & Jigsaw & 0.738 & 0.756 & 0.777 & 4.267 \\

    Swin UNETR & Masking & 0.732 & 0.763 & 0.764 & 4.756 \\
    
\end{tabular}}

\end{tabular}
\label{tab:finetune_results}
\end{table}

\begin{figure}[t!]
\centerline{\includegraphics[width=\columnwidth]{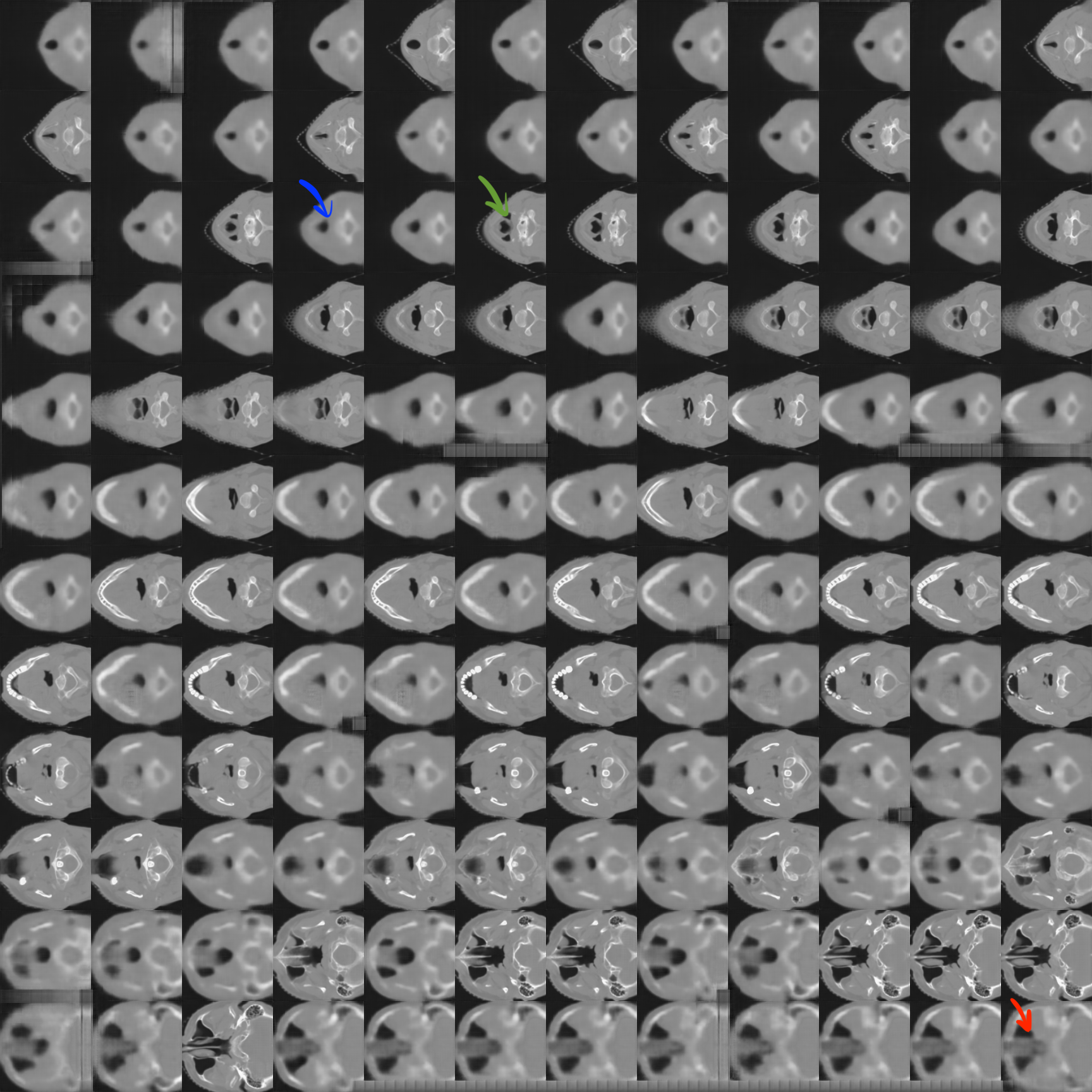}}
\caption{The figure shows the pretraining output sample for SE-norm U-Net model for the masking approach. The reconstruction seems visually reasonable, with a few errors in the masked regions. The red arrow shows a masked region that is poorly reconstructed; the blue arrow shows a well reconstructed masked region; and the green arrow shows a region that was not masked. It can be assumed that the model learnt relatively useful features during pretraining for the reconstruction (e.g. masking) task that should be helpful during finetuning.}
\label{masking_pretraining}
\end{figure}

Figure~\ref{masking_pretraining} illustrates a sample output for the pretrained model reconstruction using the SE-norm U-Net. The reconstruction is visually reasonable, with only a few mistakes in the masked regions. Three regions are highlighted: red indicates a masked region that is poorly restored, blue shows a masked region that is well restored, and green shows a non-masked region that was reconstructed well during pretraining. We assume that the model learnt useful features during pretraining for the masking task that should help the finetuning performance.

\section{Discussion}
The proposed concept of casting the 3D problem to 2D was validated using two datasets of different modalities and tissue types. We can see that 2D models based on SIs can achieve up-to-par or better results compared to 3D models. The head and neck tumor task is challenging and the variability in the CT and PET image appearance is large. When visually analyzed, we found that CT scans in this dataset for underperforming models contain artifacts, generally in the teeth area. However, when we investigated the segmentation results on multiple scans, the 2D model on SIs performs well with tumor edges, generally over segmenting, whereas the 3D model ignores these tiny regions, as is exemplified in Figure~\ref{qual}. The 3D model performs better overall due to the tumor region's main areas, where it delineates the regions more accurately.

The atrial segmentation task, being completely dissimilar from the other task, did not pose as much difficulty as the first dataset. Applying basic preprocessing techniques that easily push the SI-based 2D network to have an almost 9 percent increase in DSC is a good indicator of how much it can improve.

Thoroughly analyzing the problem, we put forth four main arguments as to why it is preferred to use 2D with SIs over 3D networks. First, although sometimes that extra bit of improvement in the performance is considered useful, it is more imperative to have a feasibly deployable models when it comes to the applications. With our approach of converting the 3D data into 2D SIs, the performance in DSC is still competitive while lowering the complexity by threefold. Second, it allows to apply pretrained weights from large-scale natural image datasets on medical images. Its effect using ImageNet1k can be seen in our results. Pretraining the model on large-scale natural image datasets and finetuning it on medical applications can be much easier with this approach. Third, because of natural images and a larger general computer vision community, a higher number of 2D networks are available. Such networks first come into the 2D world before moving to the medical imaging tasks. When dealing with volumetric data using SIs, employing these models is much easier. Finally, SSL on 2D datasets and 2D networks is much simpler and quicker. Plus, the availability of SSL pretrained models is higher in the 2D community. We experimented with two methods that indeed showed improvement in DSC, but more importantly, the implementation in 2D is preferred due to the simplicity and cheaper cost.

\section{Conclusion}
In this work, we present a 2D DL method which can efficiently process 3D data, reducing the model complexity by around three times  and still reaching similar or even better performance. In the HECKTOR dataset, simple 2D models on SIs can achieve results comparable to more powerful 3D U-Net results with much less complexity. Similarly, in the atrial segmentation dataset, the approach shows promising potential, primarily when it is powered with additional preprocessing techniques. We believe there is a potential for this new method of handling the 3D medical data to reach the state-of-the-art with much less complexity, making it easier to deploy the models in real world medical applications.


%
%
%
%
\bibliographystyle{splncs04}
\bibliography{bib}




\end{document}